\documentclass[manuscript]{aastex}

\usepackage{natbib}
\usepackage{graphicx,epsf}

\def\bv {\mathbf{v}}

\def\br {\mathbf{r}}

\def\bzpm {\mathbf{z^\pm}}

\shorttitle{Self-consistent PDF of solar wind turbulent fluctuations}
\shortauthors{Sorriso-Valvo et al.}

\begin{document}

\title{Self-consistent Castaing distribution of solar wind turbulent fluctuations}

\author{
L. Sorriso-Valvo\altaffilmark{1}, 
R. Marino\altaffilmark{1,2,3}, 
L. Lijoi\altaffilmark{3}, 
S. Perri\altaffilmark{3}, 
V. Carbone\altaffilmark{3}
}

\altaffiltext{1}{NanoTec-CNR, U.O.S. di Cosenza, Ponte P. Bucci, cubo 31C, 87036 Rende, Italy.}
\altaffiltext{2}{Space Sciences Laboratory, University of California, 7 Gauss way, Berkeley 94720, California, USA.}
\altaffiltext{3}{Dipartimento di Fisica, Universit\`a della Calabria, Ponte P. Bucci, cubo 31C, 87036 Rende, Italy.}

\email{sorriso@fis.unical.it}

\begin{abstract}
The intermittent behavior of solar wind turbulent fluctuations has often been investigated through the modeling of their probability distribution functions (PDFs). Among others, the Castaing model~\citep{castaing} has successfully been used in the past. In this paper, the energy dissipation field of solar wind turbulence has been studied for fast, slow and polar wind samples recorded by Helios 2 and Ulysses spacecraft. The statistical description of the dissipation rate has then be used to remove intermittency through conditioning of the PDFs. Based on such observation, a self-consistent, parameter-free Castaing model is presented. The self-consistent model is tested against experimental PDFs, showing good agreement and supporting the picture of a multifractal energy cascade at the origin of solar wind intermittency.
\end{abstract}




\section{Introduction}
\label{intro}

The interplanetary space is permeated by the solar wind, a rarefied, magnetized plasma continuously expanding from the solar corona. The solar wind blows radially away from the sun, and extends up to about 100AU, at supersonic and superalfv\'enic speed. Measurements collected by spacecraft instruments during last decades have evidenced that low frequency fluctuations have power law spectra~\citep{coleman}. This supports the study of fluctuations in the framework of magnetohydrodynamic (MHD) turbulence~\citep{living,biskamp}. The MHD nature of the turbulent fluctuations has recently been confirmed by more detailed analysis of the linear scaling of the mixed third order moment~\citep{polpo,forman,mcbride,prl,comment,apjl,marino_11,marino_12,prlcomp}. 
One of the most interesting properties of solar wind turbulence is the intermittent character of the fluctuations~\citep{frisch} of fields such as velocity, magnetic field, or the Elsasser fields. Intermittency is related to the non-homogeneous generation of energetic structures in the flow, due to the nonlinear transfer of energy across the scales as observed in geophysical flows \citep{pouquet_13prl,marino_13epl,marino_14pre,marino_15prl} and heliospheric plasmas \citep{living}, whose efficiency can be correlated with levels of cross-helicity \citep{smith} or self-generated kinetic helicity \citep{marino_13pre}.
Being ubiquitous in turbulence, intermittency plays a relevant role in the statistical description of the field fluctuations. The main manifestation of intermittency in fully developed turbulence is the scale dependent variation of the statistical properties of the field increments, customarily defined as $\Delta\psi_\ell(x)=\psi(x+\ell)-\psi(x)$ for a unidimensional generic field $\psi(x)$ at the scale $\ell$. In particular, many studies have focused on scaling properties of the probability distribution functions (PDFs) of the field fluctuations, $P(\Delta\psi_\ell$), and of the structure functions, defined as the moments of the distribution function of the field fluctuations $S_p(\Delta\psi_\ell)=\langle\Delta\psi_\ell(x)^p\rangle$ (here $\langle\cdot\rangle$ indicates an ensemble average). The intermittent (i.e. non-homogeneous) concentration of turbulent energy on small scale structures, as the energy is transferred through the scales, results in the enhancement of the PDFs tails, indicating that large amplitude fluctuations are more and more probable at smaller and smaller scale. Correspondingly, the structure functions scaling exponents deviate from the linear prescription valid for scale invariant PDF~\citep{frisch}. Such deviation has been studied since early works on turbulence. 
Methods based on wavelet transform~\citep{farge,onorato} or threshold techniques~\citep{greco} have been developed for the identification, description and characterization of the intermittent structures.
On the other hand, models for intermittency try to reproduce the shape of the PDFs~\citep{castaing,vergassola,edt,frisch} or the structure functions anomaly~\citep{p,she}, allowing the quantitative evaluation of intermittency.

In solar wind plasma, intermittency has been extensively characetrized in the recent years~\citep{marschtu,marsch97,sorriso99,sorriso01,chap,chapman,voros,macek,Kiyani,ragot}. As result of intermittent processes, small scale structures, like for example thin current sheets or tangential discontinuities, have been identified in solar wind~\citep{veltri,greco,bruno04,damicis}. Their dependence on parameters such as solar activity level, heliocentric distance, heliolatitude and wind speed has been also pointed out~\citep{pagel02,radial,chapman-radial,chapman-active,yorda}. 

One of the models for the description of the increments PDFs was introduced by Castaing~\citep{castaing}, and successfully applied in several contexts~\citep{sorriso99,sorriso2d,sorriso01,padhye,pagel,stepanova,chapman,leubner,voros,cimento}. 
Within the multifractal framework~\citep{frisch}, for each scale, the energy transfer rate $\epsilon_\ell$ has non-homogeneous scaling properties, as for example the fractal dimension (which is directly related to the cascade efficiency) in different regions of space. The turbulent fields can be thus interpreted as a superposition of subsets, each characterized by a given fractal dimension, and with a typical energy transfer rate. Each region can then be reasonably assumed to have the same distribution $P_0$ of the field fluctuations, with variable width $\sigma$ (depending on the cascade efficiency, and related with the local fractal dimension) and weight $L(\sigma)$ (depending on the fraction of space characterized by the same statistics). 
The Castaing model PDF consists thus of the continuous superposition of such distributions, each contributing to the statistics with its appropriate weight. The latter is introduced through the distribution function of the widths $\sigma$~\citep{cimento}. This leads, for each time scale $\ell$, to the convolution:
  \begin{equation}
   P(\Delta\psi_\ell)=\int_0^\infty L_\lambda(\sigma_\ell)P_0(\Delta\psi_\ell,\sigma_\ell) \; d \ \ln\sigma_\ell \, .
  \label{convolution}
  \end{equation}
Based on empirical large scale PDF shape, a Gaussian parent distribution is normally used. 
It is known from turbulence studies that PDFs of fluctuations have to be skewed. Indeed, symmetric PDFs would result in vanishing odd-order moments of the fluctuations, in contrast with experimental results~\citep{frisch}. Asymmetric PDFs are also necessary to satisfy the linear scaling of the (non-vanishing) third-order moment of the fluctuations, as required by theoretical results~\citep{frisch,polpo}. Thus, in order to account for this, a skewness parameter $a_s$ must also be included, so that 
  \begin{equation}
 P_0(\Delta\psi_\ell,\sigma_\ell,a_s) = \frac{1}{\sqrt{2\pi\sigma} } \exp \left[ -\frac{\Delta\psi_\ell^2}{2\sigma^2} \left( 1+a_s \frac{\Delta\psi_\ell/\sigma}{(1+\Delta\psi_\ell^2/\sigma^2)^{1/2}} \right) \right] \, .
  \label{gaussian}
  \end{equation}
The PDF of variances $L_\lambda(\sigma_\ell)$ needs theoretical prescription. A log-normal ansatz has been often used, as conjectured in the framework of the multifractal cascade~\citep{castaing}
\begin{equation}
L_{\lambda(\sigma_\ell)} = \frac{1}{\lambda_\ell\sqrt{2\pi}} \exp \left[ \frac{\ln^2\sigma_\ell/\sigma_{0,\br}}{2\lambda^2_\ell} \right].
\label{lognorm}
\end{equation}
Such choice has been justified by assuming that the nonlinear energy transfer is the result of a multifractal fragmentation process, giving rise to a multiplicative hierarchy of energetic structures. By assuming random distribution of the multipliers (namely, of the local efficiency of the cascade), the central limit theorem suggests a log-normal distribution of the local energy transfer rate. Then, from dimensional considerations, the fluctuations variance can be expected to share the same statistical properties as the energy transfer rate, therefore giving the log-normal distribution~(\ref{lognorm})~\citep{castaing}. 
In equation (\ref{lognorm}), for $\lambda^2_\ell=0$, the log-normal PDF is a $\delta$-function, so that the convolution (\ref{convolution}) gives one Gaussian of width $\sigma_{0,\ell}$ the most probable value of $\sigma_\ell$. As $\lambda^2_\ell$ increases, the convolution includes more and more values of $\sigma_\ell$, and the PDF tails are enhanced.
Therefore, the scaling of the parameter $\lambda^2_\ell$ controls the shape of the PDF tails, and describes the deviation from the parent distribution, characterizing the intermittency in the inertial range~\citep{castaing,sorriso99}. In fully developed turbulence, a power-law scaling is usually observed for  $\lambda^2_\ell$~\citep{castaing,sorriso99}. Finally, a relationship can be established between the scaling exponent of $\lambda^2_\ell$ and the multifractal properties of the flow~\citep{castaing}. 

As briefly described above, the Castaing model is based on hypotheses on the physical processes governing the turbulent cascade. The main assumption is the choice of the weights distribution, which needs appropriate theoretical modeling. In this paper, after verifying that the multifractal cascade framework applies to solar wind turbulence, we show that it is possible to describe solar wind intermittency without any hypothesis on the shape of such distribution, but rather using empirical weights~\citep{naert}. In Section~\ref{data} we briefly introduce the data used for the analysis; in Section~\ref{epsilon} we estimate the empirical distribution function of the local energy dissipation rate; Section~\ref{conditioned} shows the conditioned analysis performed on the data in order to extract the empirical weights for the Castaing model, the self-consistent Castaing probability distribution functions are built and compared with the experimental PDFs.

\section{The data: Helios 2 and Ulysses}
\label{data}

This work is based on the analysis of three different samples of {\it in situ} measurements of velocity ${\bf v}$, mass density $\rho$, estimated using proton and $\alpha$ particle, and magnetic field ${\bf B}$. The Elsasser variables ${\bf z}^\pm={\bf v}\pm {\bf B}/\sqrt{4\pi\rho}$ have also been evaluated from the time series. 
Two samples were taken in the ecliptic wind by the Helios 2 spacecraft during the first 4 months of 1976, when the spacecraft orbited from $1$AU on day 17, to $0.29$AU on day 108. Data resolution is $81$ seconds, and eleven fast or slow wind streams, each about $49$ hours long, have been extracted to avoid stream-stream interfaces and to ensure a better stationarity~\citep{sorriso99}. As pointed out in the literature, fast and slow wind turbulence should be studied separately, because of the different plasma conditions~\citep{living}. Therefore, for our statistical analysis we built two distinct samples by putting together six streams for the fast wind (which we name the {\it fast} sample, totalizing $N^F_{tot}=12288$ data points), and five streams for the slow wind (hereafter the {\it slow} sample, consisting of $N^S_{tot}=10240$ data points).
The third sample was recorded in the solar wind out of the ecliptic, by instruments on-board Ulysses spacecraft, during the first eight months of 1996. The spacecraft spanned distances in the range from $3.1$AU to $4.7$AU, and latitudes from about $53^\circ$ to $19^\circ$. Sampling resolution is $8$ minutes. We refer to this dataset as {\it polar} sample.
All samples were taken at solar minima, when the solar wind is more steady, and free from disturbances of solar origin.
We remind that, in order to study spacecraft time series, all scales $\ell$ are customarily transformed in the time lags $\tau = \ell / |\langle \bv \rangle|$ through the bulk flow speed averaged over the entire data set. This is allowed by the Taylor hypothesis, that is generally valid for solar wind fluctuations in the inertial range~\citep{perri,living}, and which we have tested in our data. 
%
%
\begin{figure}  
\begin{center}
\includegraphics[width=16cm]{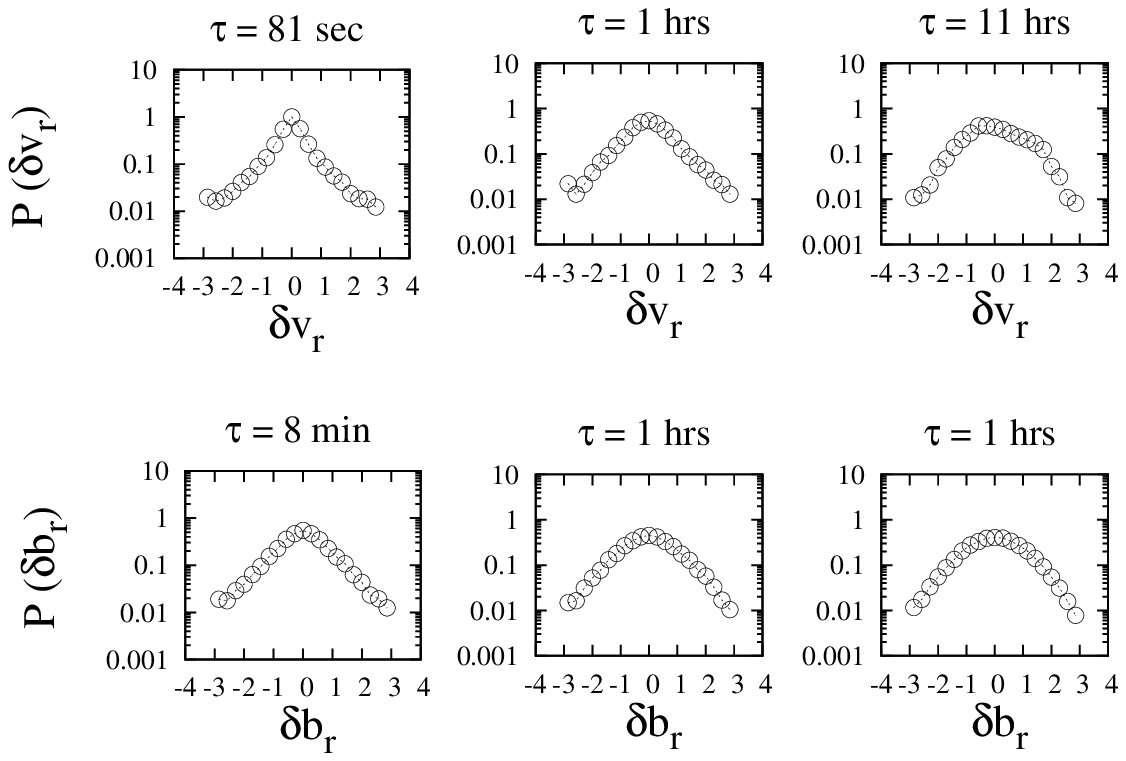}     
\caption{The PDFs of standardized field fluctuations at three different scales (increasing from left to right), for fast solar wind velocity fluctuations from Helios 2 (top panels), and for polar Ulysses magnetic fluctuations (bottom panels).}
\label{Fig-pdfs}
\end{center}
\end{figure}
In Figure~\ref{Fig-pdfs}, samples of the PDFs of standardized magnetic and velocity field fluctuations at different scales are shown for fast and polar samples~\citep{sorriso99,sorriso01,cimento,electric}. 
The typical tail enhancement toward small scales appears evident for all cases~\citep{sorriso99}. Similar results are observed for the velocity and for the Elsasser fields $\bzpm$ (see e.g.~\citet{sorriso99,cimento}). PDFs of magnetic field, velocity and Elsasser fields fluctuations have been successfully reproduced using the Castaing model given in Equation~(\ref{convolution})~\citep{sorriso99}. In particular, the heavy tails are extremely well captured by the model, showing that the effect of intermittency has been well described. The fitting parameter $\lambda^2$ shows a power-law scaling extended over about one decade. As mentioned above, the scaling exponents are related to the fractal dimension of the most intermittent structures generated at the bottom of the cascade~\citep{castaing}, therefore providing important physical information on the turbulent flow properties.

\section{Statistical properties of the energy dissipation rate}
\label{epsilon}

\begin{figure}
\begin{center}
\includegraphics[width=16cm]{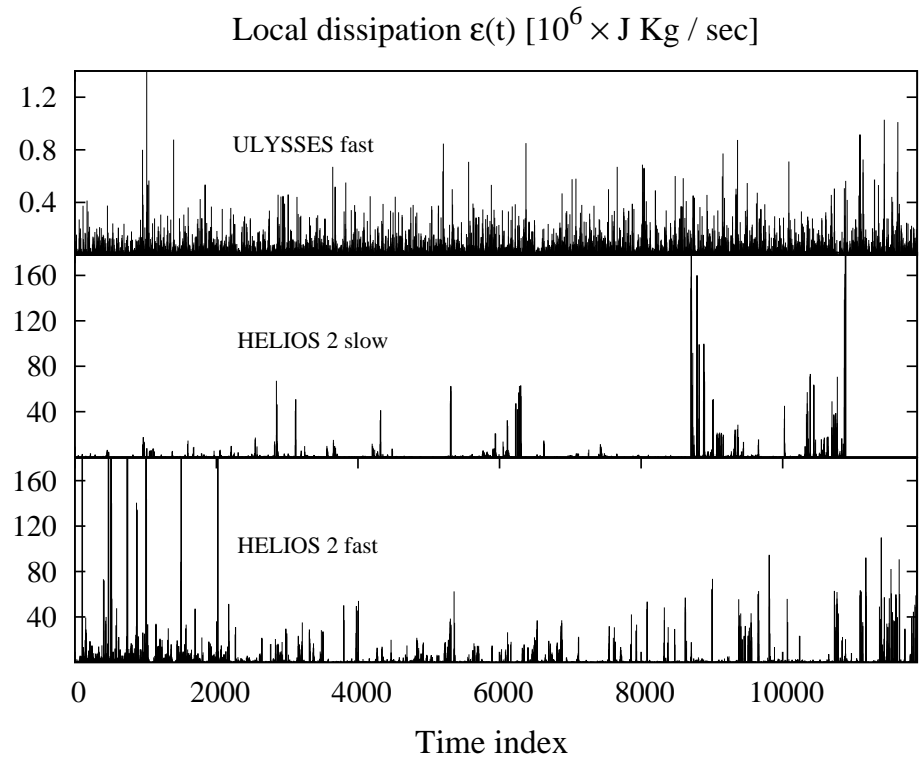}
\caption{The local energy dissipation rate (see text) for the three datasets:  Ulysses fast polar wind (top), Helios 2 fast streams (center), and Helios 2 slow streams (bottom).} 
\label{Fig-epsilon}
\end{center}
\end{figure}

\begin{figure}
\begin{center}
\includegraphics[width=10cm]{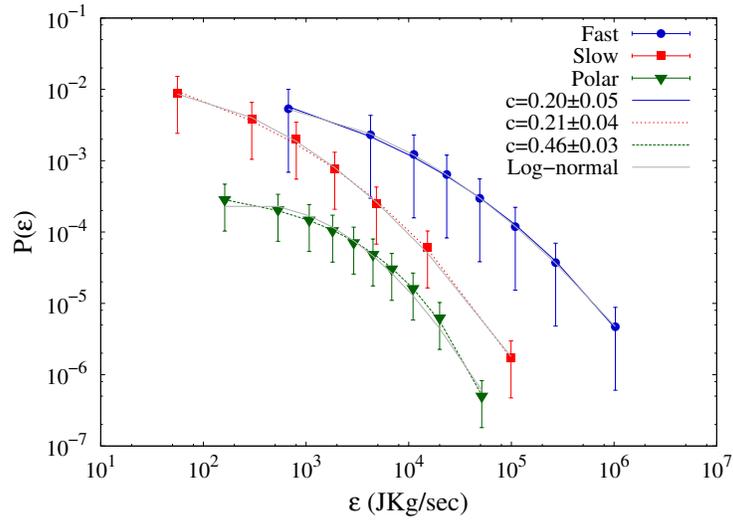}  
 \caption{The probability distribution function of the energy dissipation rate $\epsilon$ estimated from the Helios 2 data for fast wind (blue circles) and slow wind (red squares), and from the Ulysses polar data (green triangles). Fits with log-normal (grey lines) and with a stretched exponential function (see legend) are superposed, and the corresponding shape exponents $c$ are indicated. The three distributions have been vertically shifted for clarity.}
\label{fig-pepsi}
\end{center}	
\end{figure}

The first step for the self-consistent characterization of the PDFs is to map the properties of the plasma energy transfer rate using the experimental time series. In solar wind turbulence, understanding the mechanism responsible for the dissipation (and/or dispersion) of the energy at the bottom of the nonlinear cascade is still an open issue. Therefore, the actual expression of the dissipative (and/or dispersive) terms within MHD equations is not known. 
It should also be pointed out that, unlike ordinary turbulent flows, the solar wind plasma is only weakly collisional, so that molecular viscosity and resistivity cannot be defined in a simple way, nor estimated directly from the measurements. These considerations show that it is not possible, to date, to measure the local energy dissipation rate in solar wind turbulence. 
However, it is possible to define proxies of the energy dissipation rate, that can be reasonably used to represent the statistical properties of the field.
In this paper, we use a definition based on the third order moment scaling law for MHD~\citep{polpo,prl}, often referred to as Politano-Pouquet law (PP). The PP law establishes, under given hypotheses (stationarity, homogeneity, isotropy, incompressibility), the linear scaling of the mixed third order moment of the Elsasser fields, 
\begin{equation}
    Y^\pm(\tau) = \left \langle |\Delta \bzpm_\tau(t)|^2\, \Delta z^\mp_\tau(t)\right\rangle = -\frac{4}{3} \,\langle\epsilon^\pm\rangle\tau \langle v \rangle \; .
\label{yaglom}
\end{equation}
In the right hand side of Equation~(\ref{yaglom}), $\langle\epsilon^\pm\rangle$ is the mean energy transfer rate, estimated over the whole domain. By analogy, we define the ``local'' pseudo-energy transfer rate as: 
\begin{equation}
    \epsilon^\pm_\tau (t) = \frac{|\Delta z^\pm(t)|^2\Delta z^\mp_R(t)}{\tau \langle v \rangle} \, ,
\label{pseudoenergy}
\end{equation}
so that the local energy transfer rate at the scale $\tau$ reads $\epsilon_\tau(t) = (\epsilon^+_\tau (t)+ \epsilon^-_\tau (t))/2$. At a given scale, each field increment can thus be associated with the local value of $\epsilon_\tau(t)$~\citep{marsch97}. 
Since we are interested, in particular, in the small scale intermittent effects, from now on we will only use the resolution scale values of $\epsilon(t)$ (namely 81 seconds for Helios 2 data and 8 minutes for Ulysses data), which we will refer to as energy dissipation rate. 
Figure~\ref{Fig-epsilon} shows the variable $\epsilon(t)$, computed for fast and slow Helios 2 streams and for the Ulysses polar sample. 
Differences between the three samples are evident, in particular for the Ulysses dataset, probably because of the lower data resolution. 
For all cases, the field is highly irregular and inhomogeneous, with spikes of large dissipation alternated with quiet periods. 
%
%
The probability distribution functions of $\epsilon(t)$ are shown in Figure~\ref{fig-pepsi} for fast and slow wind, as obtained from Helios 2 data, and for polar Ulysses data. Because of the inhomogeneity, PDFs have been computed using bins of variable width, by imposing a fixed number of data points in each bin ($N_p=1500$ for Helios datasets, $N_p=3000$ for Ulysses data). The error bars are estimated as the counting (Poisson) error on $N_p$, and the three PDFs have been vertically shifted for clarity.
Log-Normal fits of the distributions are overplotted in light grey, showing good agreement with the framework of a multiplicative cascade~\citep{castaing}. Alternatively, PDFs are even better reproduced (with two to ten times smaller $\chi^2$) by a stretched exponential fit $P(\epsilon)\sim \exp{-(\epsilon/\epsilon_0)^c}$, where $\epsilon_0$ is a characteristic value of the energy dissipation rate, and $c$ is the parameter controlling the shape of the tails of the PDF. In particular, $c=2$ indicates Gaussian, $c=1$ exponential, and $c<1$ indicates heavy tailed, almost power-law distribution. In the present case, the shape parameters are indicated in Figure~\ref{fig-pepsi}. Both fast and slow wind measured by Helios 2 show a strong deviation from Gaussian ($c\simeq0.2$), with presence of energy dissipation bursts that result in the heavy tails of the PDFs. On the contrary, Ulysses shows ``thinner'' tails, i.e. populated with less extreme events. This is to be expected, because of the different time resolution of the data. Indeed, while Ulysses sampling time sits in the middle of the inertial range, for Helios 2 it is closer to its bottom.
It should be recalled that stretched exponential PDFs were predicted for a variable generated as the result of a multiplicative process controlled by a few extreme events, in the framework of the extreme deviation theory~\citep{edt}. This supports once more the presence of a intermittent, multiplicative energy cascade in solar wind turbulence.

\section{The conditioned analysis and a self-consistent Castaing model}
\label{conditioned}

\begin{figure}
\begin{center}
 \includegraphics[width=7.5cm]{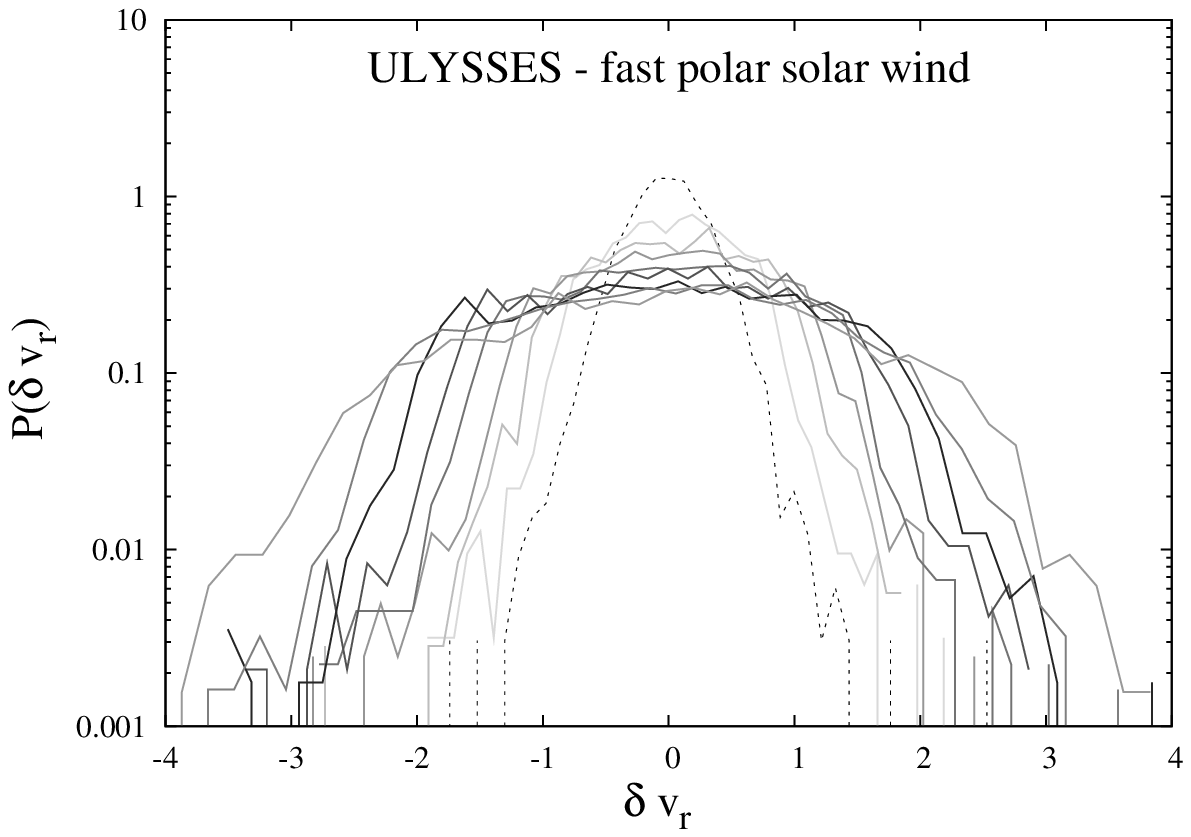}\includegraphics[width=7.5cm]{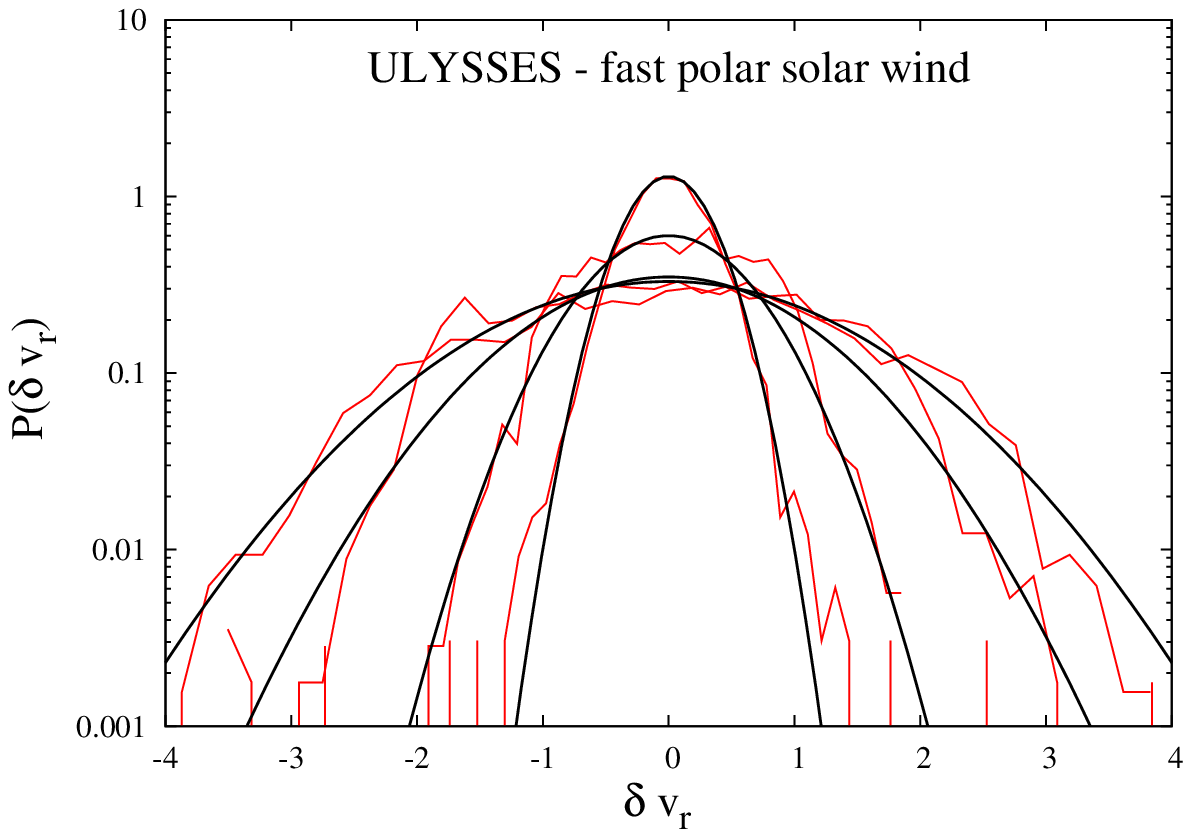} 
 \includegraphics[width=7.5cm]{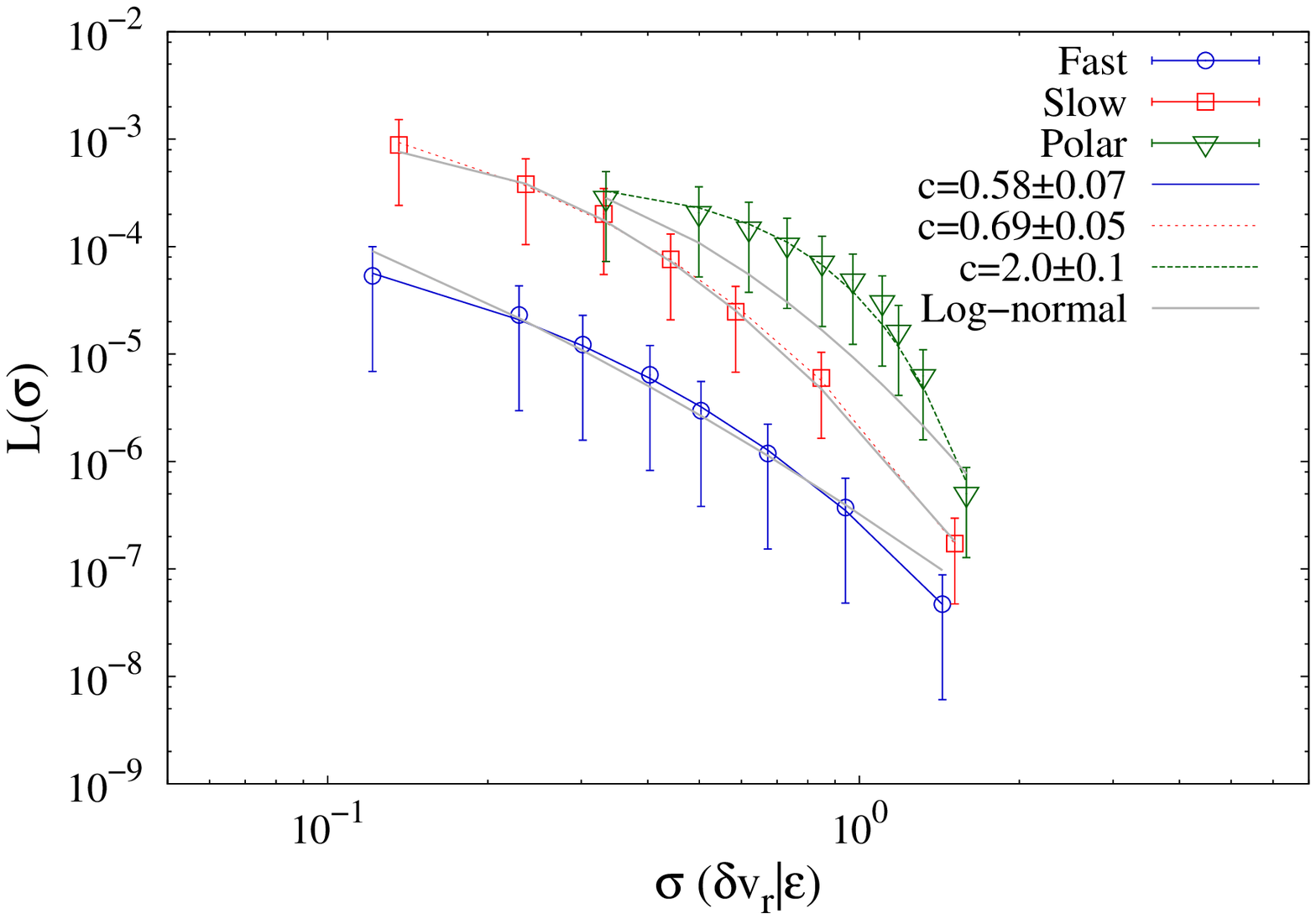}\includegraphics[width=7.5cm]{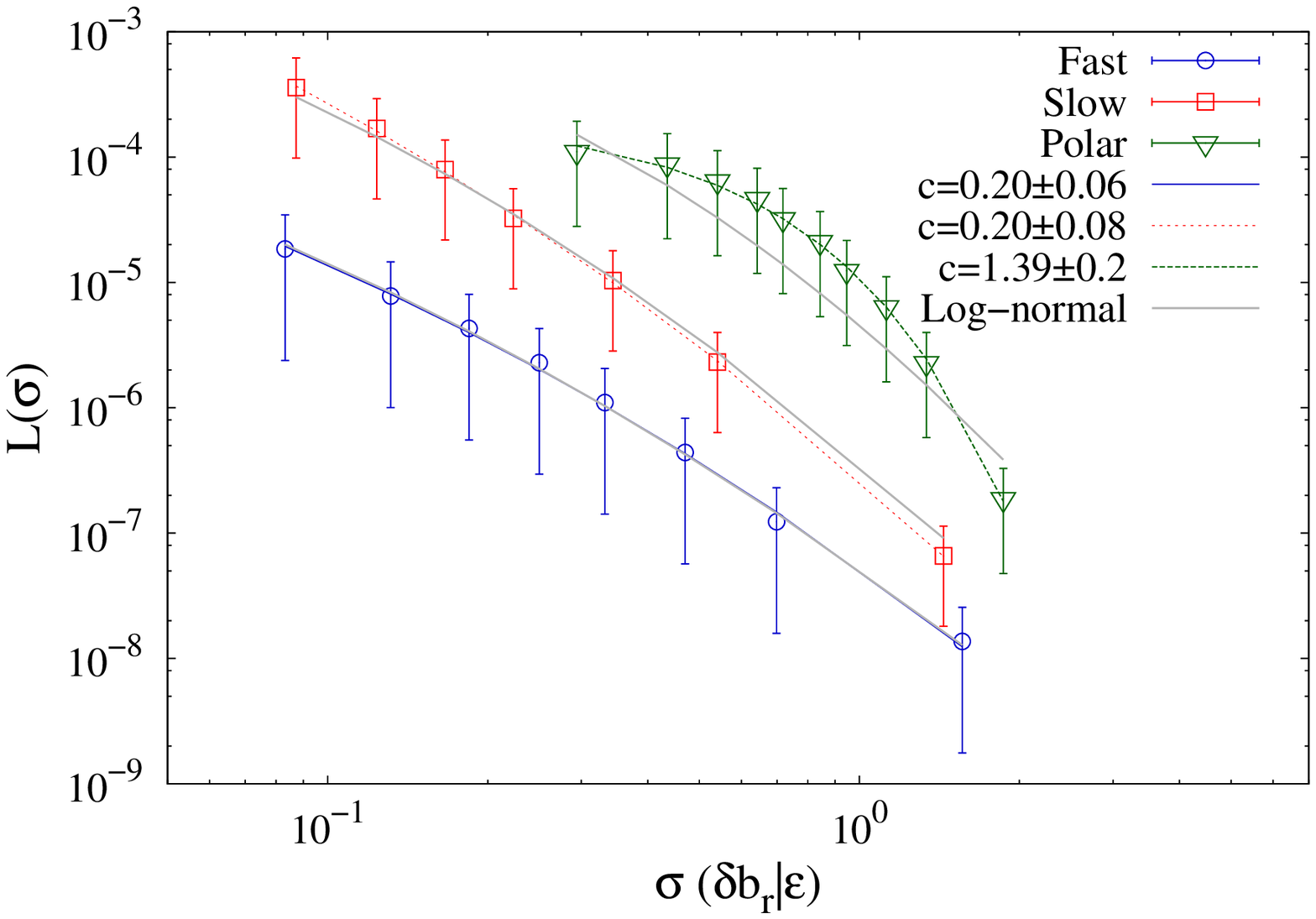} 
 \includegraphics[width=7.5cm]{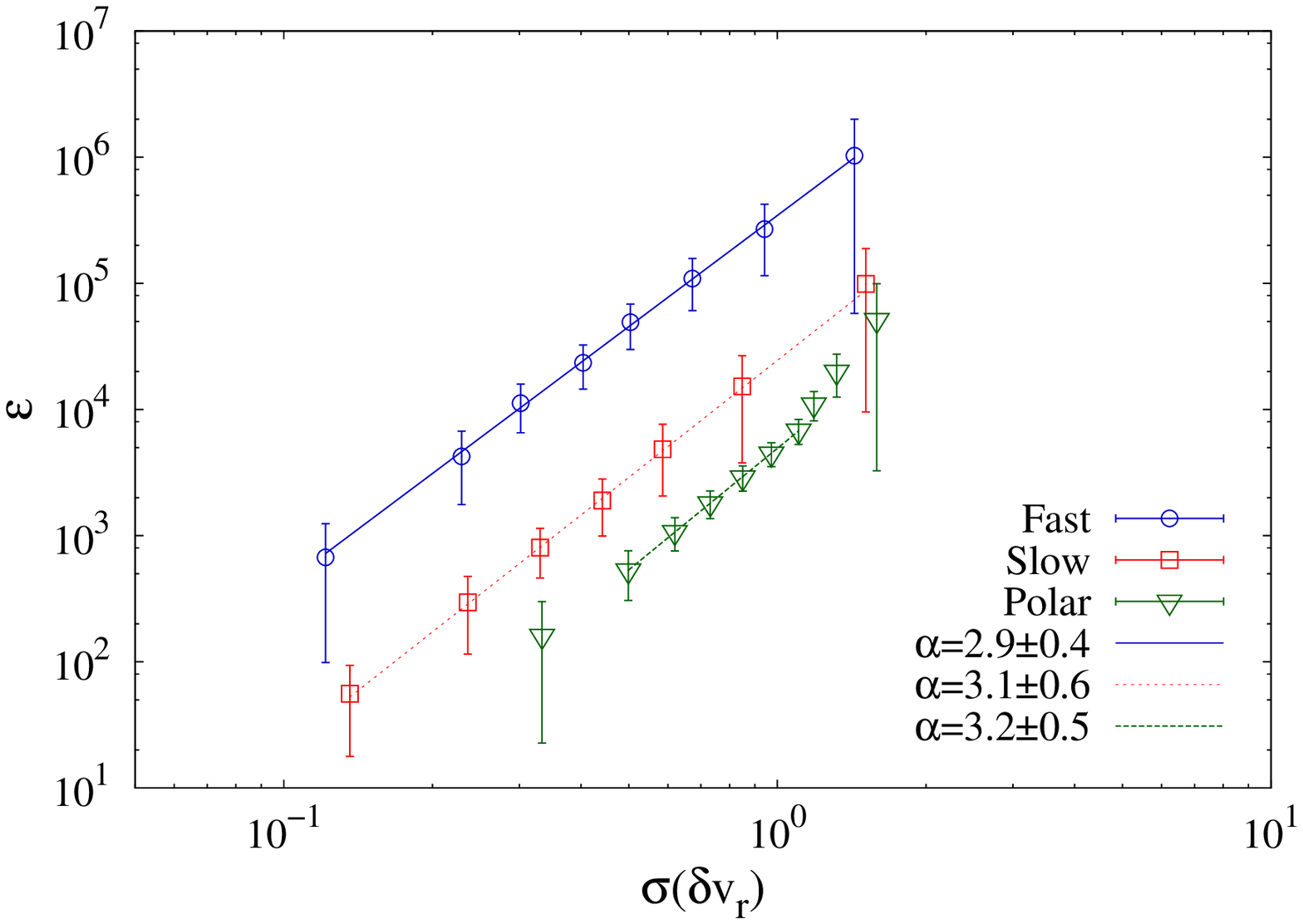}\includegraphics[width=7.5cm]{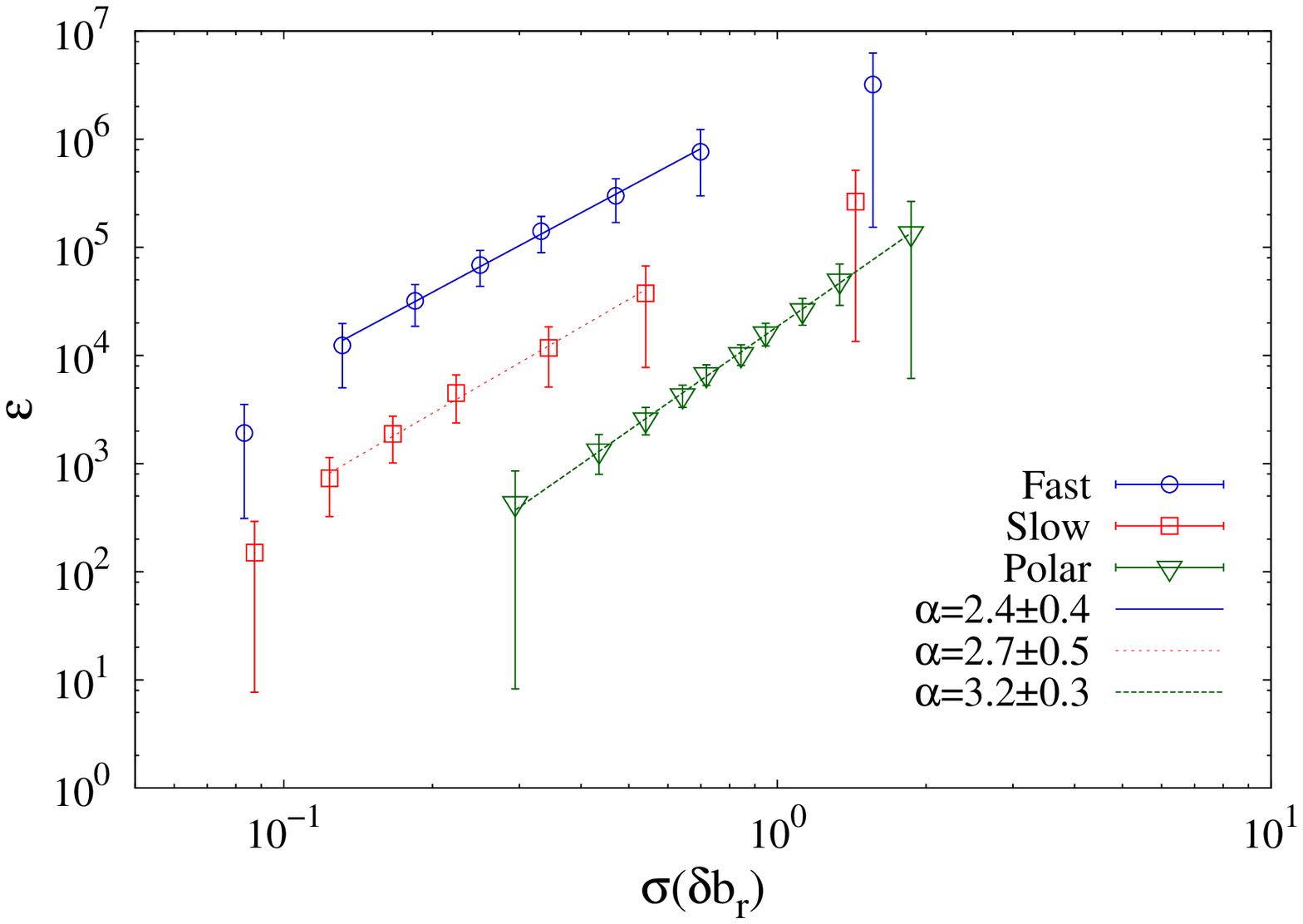} 
\end{center}
 \caption{
Top-left panel: conditioned PDFs $P(\Delta\psi|\epsilon)$ for the radial component of the velocity in the polar Ulysses sample, for ten different values of $\epsilon$. Top-right panel: the Gaussian fit of $P(\Delta\psi|\epsilon)$ for the same case, only shown for four values of $\epsilon$.
Central panels: PDFs of $\sigma(\Delta\psi|\epsilon)$ for the three dataset, for the radial component of velocity (center-left panel) and magnetic field (center-right panel). Fits with a stretched exponential law are superposed (solid lines), and the fitting parameters $c$ are indicated. 
Bottom panels: relationship between the energy dissipation rate $\epsilon$ and the variance $\sigma(\Delta\psi|\epsilon)$ for velocity (bottom-left panel) and magnetic field (bottom-right panel) fluctuations.}
\label{fig-conditioned}
\end{figure}

In order to verify that the Castaing model can properly be applied to solar wind turbulence, a conditioned analysis of the data was performed. The range of values of the energy dissipation rate was divided into $N_{bin}$ bins of variable width $A_{\epsilon}$ (with $N_{bin}=8,7,10$ for fast, slow and polar samples, respectively), in order to separate different regions of the time series. The PDFs of field fluctuations, at the resolution time lag, were thus estimated separately for each bin, i.e. conditioned to the values of the energy dissipation rate. Top-left panel of Figure~\ref{fig-conditioned} shows, superimposed, all the ten conditioned PDFs $P(\Delta\psi|\epsilon)$ of the resolution scale radial velocity field fluctuations for the polar wind dataset. Conditioned PDFs no longer have the heavy tails observed at small time lags (see Figure~\ref{Fig-pdfs}), and present roughly Gaussian shape. The top-right panel of the same Figure shows examples of the Gaussian fits of the conditioned distributions. For clarity, only four out of ten conditioning values of $\epsilon$ are plotted. This observation confirms the multifractal scenario, in which regions with the same energy dissipation rate are characterized by self-similar (Gaussian in this case) PDFs of the field fluctuations, even at small scale. Thus, the conditioned data lose intermittency, which indeed arises from the local fluctuations of the energy dissipation rate. Top panels of Figure~\ref{fig-conditioned} also qualitatively highlights the result of superimposing several Gaussian distributions of variable width and amplitude, obtaining the heavy tails observed in the usual PDF. The use of the Castaing distribution is therefore appropriate to describe solar wind intermittent turbulence.
For each conditioned PDF, corresponding to a given set of constant dissipation rate, it is possible to obtain the probability density function of the standard deviations $L(\sigma)=N_p/N_{tot}A_{\epsilon}$, where $N_p$ is the (fixed) number of points in each dissipation bin. This is done for each dataset by exploiting the correspondence between a given bin of energy dissipation rate $\epsilon$ and the standard deviation of the corresponding field fluctuations. In other words, for each field $\psi$ and for each value of $\epsilon$, a value of $\sigma(\Delta\psi|\epsilon)$ is obtained through the Gaussian fit of the conditioned PDFs $P(\Delta\psi|\epsilon)$. Then, the probability $L(\sigma)$ is the fraction of data associated with that interval of $\epsilon$ values. This allows to establish an explicit relationship between the statistical properties of the energy dissipation rate $\epsilon$ and of the standard deviation $\sigma$, postulated in the Castaing model. Central panels of Figure~\ref{fig-conditioned} shows the probability distribution functions of the standard deviation of the fields $L(\sigma)$ associated to the distributions of $\epsilon$, along with log-normal and stretched exponential fits. In this case, the log-normal fit is not always able to capture the form of the distribution, while the stretched exponential fits are very good. For the latter, the smaller exponents $c$ observed for the velocity fluctuations in all dataset, and suggesting a lower intermittency, agree with the higher level of intermittency of the magnetic field~\citep{sorriso99}. It is also interesting to check directly from the data the validity of the assumption originally used in the Castaing model, namely the phenomenological relationship between the energy dissipation rate and the variance of the fluctuations. This can be simply done by plotting $\epsilon$ {\it versus} $\sigma(\Delta\psi|\epsilon)$, and verifying the power-law relation $\epsilon \sim \sigma^\alpha$ expected from a simple dimensional argument on the third orer scaling relation given by Eq.~(\ref{yaglom}), resulting in $\alpha=3$~\citep{castaing,naert}. As can be seen in the bottom panels of Figure~\ref{fig-conditioned}, the power-law relationship is very well verified, with exponents compatible with the expected value $\alpha=3$. The error bars are estimated as half the size of the energy dissipation rate bin. Similar results apply to the other components of the fields and to the Elsasser fields (not shown).
Note that a similar analysis performed on ordinary fluid turbulence provided instead $\alpha=5$~\citep{naert}.

\begin{figure}
\begin{center}
\includegraphics[width=10cm]{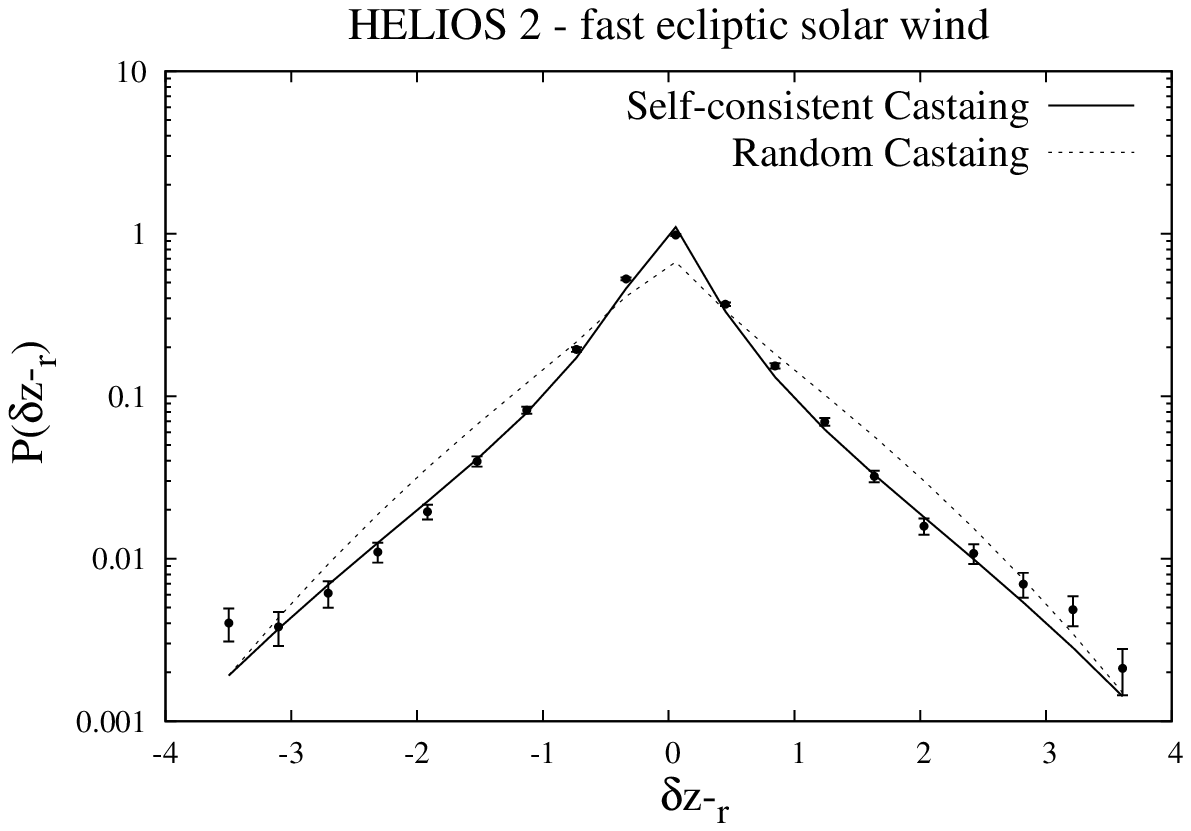}
\includegraphics[width=10cm]{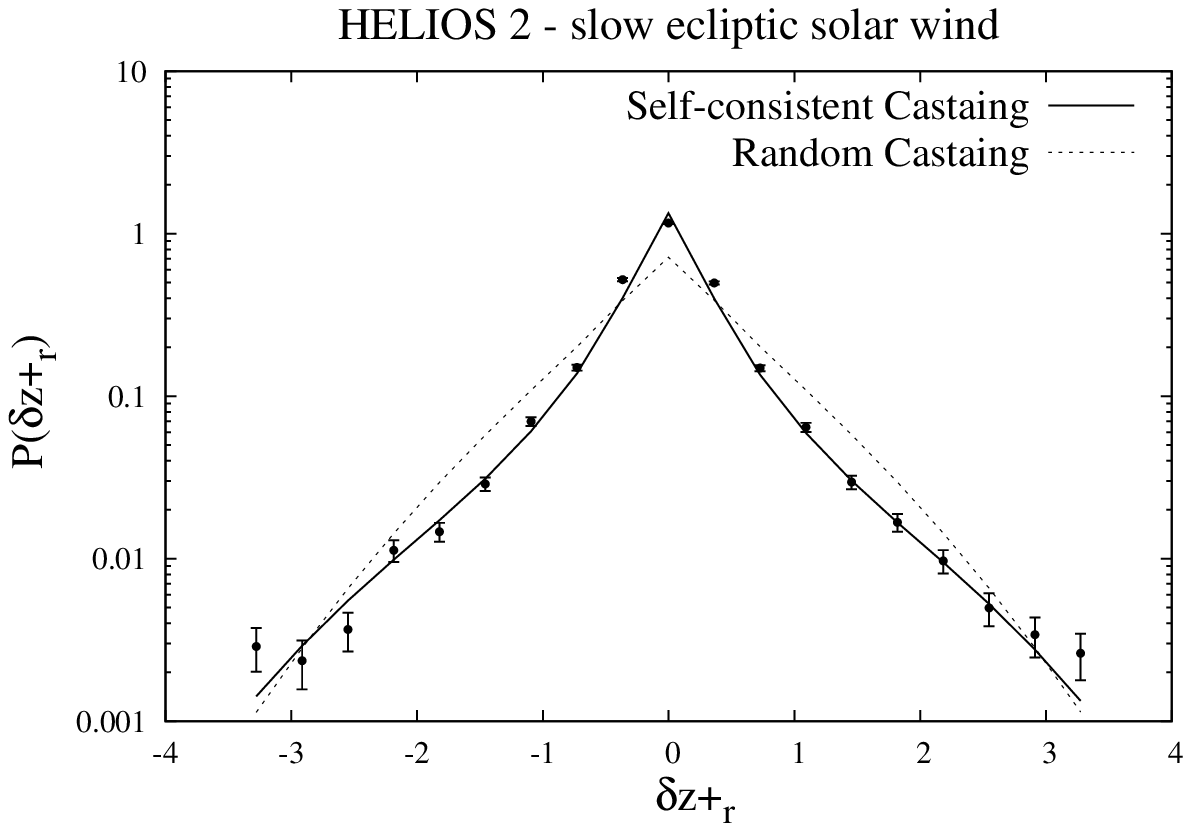}
\includegraphics[width=10cm]{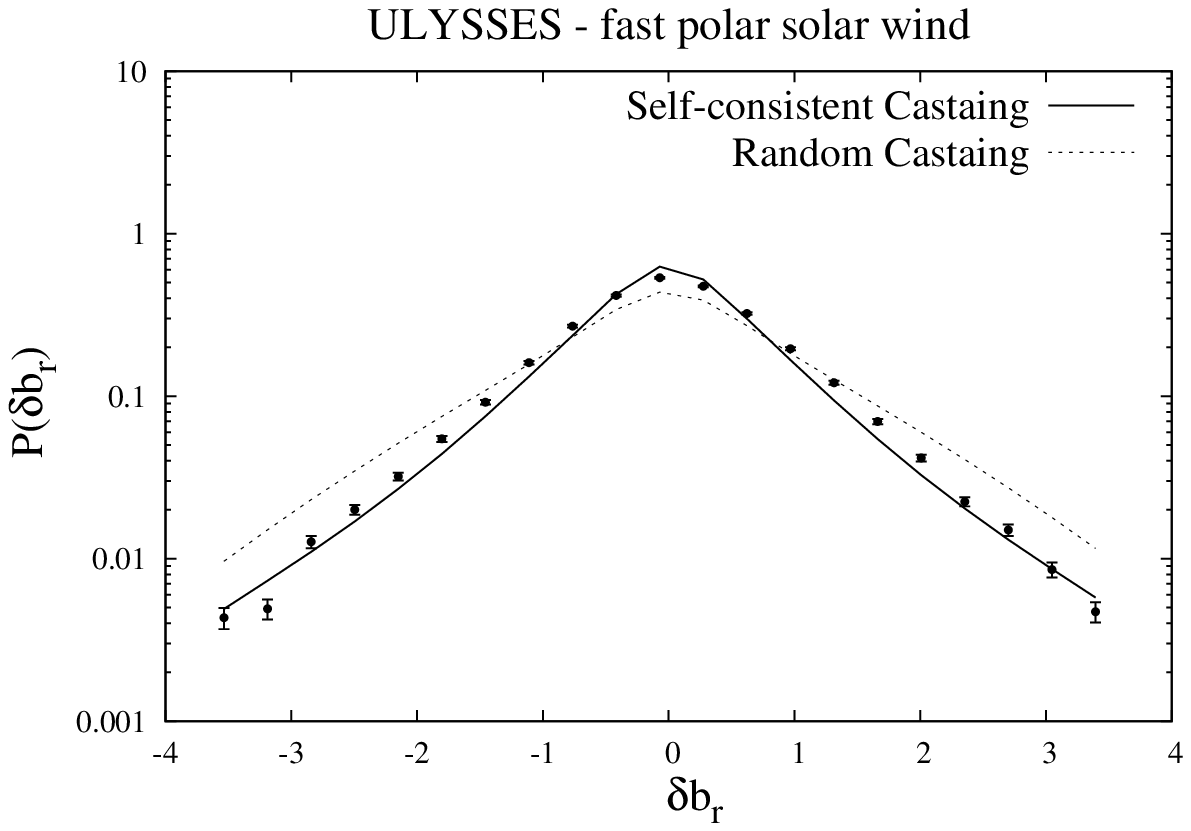}
 \caption{The PDFs of radial Elsasser or magnetic fields fluctuations at the resolution scale, computed for the three samples, together with the self-consistent Castaing model (full line). The dotted line indicates the model PDF obtained for random distribution of the variances (see text).}
\label{fig-self-castaing}
\end{center}
\end{figure}

With all this information at hands, the self-consistent Castaing distribution~(\ref{convolution}) can be finally discretized as follows: 
\begin{equation}
\label{selfconsistent}
  P(\Delta\psi_\tau)=A_{norm}\sum_{i=1}^{N_{bin}} {L(\sigma) P_0(\Delta\psi_\tau,\sigma,{a_s}) \; A_{\epsilon} }.
\end{equation}
Equation~(\ref{selfconsistent}) only includes two free parameters: the normalization factor $A_{norm}$, and the skewness factor $a_s$, while the information on the PDF shape is self-consistently included through the distribution $L(\sigma)$. 
The PDFs of the fields fluctuations can now be compared with the self-consistent Castaing model, by fitting the data with Equation~(\ref{selfconsistent}) in order to adjust only for the two free parameters. These are however irrelevant for the tails curvature, i.e. for intermittency, so that there is no need to discuss them here. 
Figure~\ref{fig-self-castaing} shows the PDFs of fluctuations, as in previous Figure~\ref{Fig-pdfs} (markers), together with the fit with Equation~(\ref{selfconsistent}) (full line). The self-consistent model reproduces the shape of the PDF tails very satisfactorily in all cases. 
In order to test the goodness of the fit, a synthetic $L_R(\sigma)$ has been built by averaging over $10^4$ realizations of random distributions $L(\sigma)$, generated for $\sigma$ in the same interval as each solar wind dataset, $[\sigma_{min}, ... , \sigma_{max}]$. The fit of the data with the self-consistent Castaing model, using the synthetic $L_R(\sigma)$, is shown in Figure~\ref{fig-self-castaing} as dotted line. It is evident that a superposition of Gaussians with a random choice of the weigth distribution does not allow the correct reproduction of the data. Quantitatively, the random Gaussian superposition fits have $\chi^2$ values larger by a factor of ten with respect to the self-consistent Castaing fit.

\section{Summary}

In the attempt to understand the mechanism responsible for the generation of the highly intermittent turbulent fluctuations of solar wind fields, we have explored here the existence of a non-homogeneous energy cascade, a base hypothesis of several models. 
The intermittent solar wind field fluctuations, deeply investigated in the last decade, have often been described through the Castaing PDF model. In this work, we have evaluated empirically the weights distribution $L(\sigma)$ of the Castaing model (Eq.~(\ref{convolution})), rather than assuming an analytical prescription (usually a log-normal distribution). 
For the sake of generality, we have selected three different datasets, consisting of one fast and one slow ecliptic wind samples (Helios 2 data), and one fast polar wind sample (Ulysses data).
Since it is not possible, to date, to measure the energy dissipation rate from the data, we have used a proxy derived from the exact statistical scaling law for the mixed third order moment of the magnetohydrodynamic fluctuations. Although such law is expected to apply only for homogeneous, isotropic, incompressible MHD turbulence, its validity in the solar wind time series has recently been established by several authors (e.g.~\citet{sorriso99}. Therefore, the use of the proxy introduced here is appropriate. 
After estimating the local value of the energy dissipation rate, we have evaluated its statistical properties. The probability distribution functions are consistent with a log-normal function, predicted by the typical theoretical models used for the description of the turbulent fields as the result of a non-homogeneous, random multiplicative cascade.
Alternatively, a stretched exponential fit was also used for all datasets. Stretched exponential distributions match the theoretical framework of the Extreme Deviations Theory (EDT)~\citep{edt}, where the statistical properties of the multiplicative cascade process are controlled by the most extreme intermittent events (the heavy tails of the PDFs). The stretching parameter, obtained from the fit, can be used to describe quantitatively the degree of inhomogeneity, being related to presence of correlations or clustering in the dissipation field. The results obtained here (see Figure~\ref{fig-pepsi}) confirm the stronger intermittency of slow wind with respect to fast and polar wind~\citep{sorriso99,living}. 

Starting from this observation, we have then studied the conditioned PDFs of the fields fluctuation, the conditioning parameter being the local energy dissipation rate. As expected within the multifractal picture (but never observed in the solar wind so far), conditioning results in loss of intermittency, so that self-similarity is restored when the inhomogeneity of the energy dissipation rate is removed. This result shows that the framework of the multifractal energy cascade applies to solar wind turbulence.

Upon observation of their Gaussian shape, the conditioned PDFs have finally been used to build the Castaing distribution self-consistently, through the empirical distribution of their standard deviations, $L(\sigma)$. The model PDF drawn following this procedure fits very well the shape of the experimental PDFs, and in particular can capture the curvature of the tails. The self-consistent Castaing model can therefore reproduce the intermittency of the fields. 
We remark that, the selection process for the conditioned PDFs being based on the dissipation rate, the relevant physical information necessary to describe the intermittent PDFs is the non-homogeneous distribution of the local energy dissipation rate. 
The role of the shape parameter $\lambda^2$, used to describe quantitatively the degree of intermittency, is now played by the stretching parameter $c$ of the distributions of the energy dissipation rate, not directly involved in the self-consistent Castaing model.

In conclusion, the stretched exponential distribution of the local energy dissipation rate, the Gaussian shape of the conditioned PDFs, and the successful application of the self-consistent Castaing model, all strongly support the picture of the multifractal energy cascade being at the origin of the intermittency of solar wind turbulence. The validity of such scenario holds for both fast, slow and polar samples, confirming that the turbulent cascade is always active in the solar wind.

\begin{acknowledgements}
The research leading to these results has received funding from the European Community’s Seventh Framework Programme (FP7/2007-2013) under grant agreement n. 269297/``TURBOPLASMAS''.
\end{acknowledgements}

{}

\end{document}